# Multi-spacecraft Analysis of the Properties of Magnetohydrodynamic Fluctuations in Sub-Alfvénic Solar Wind Turbulence at 1 AU

S. Q. Zhao[12], Huirong Yan[126], Terry Z. Liu[36], Mingzhe Liu[4], Huizi Wang[5]

**Abstract**

We present observations of three-dimensional magnetic power spectra in wavevector space to investigate the anisotropy and scalings of sub-Alfvénic solar wind turbulence at magnetohydrodynamic (MHD) scale using the Magnetospheric Multiscale spacecraft. The magnetic power distributions are organized in a new coordinate determined by wavevectors ($\hat{k}$) and background magnetic field ($\hat{b}_0$) in Fourier space. This study utilizes two approaches to determine wavevectors: the singular value decomposition method and multi-spacecraft timing analysis. The combination of the two methods allows an examination of the properties of magnetic field fluctuations in terms of mode compositions without any spatiotemporal hypothesis. Observations show that fluctuations ($\delta B_{\perp 1}$) in the direction perpendicular to $\hat{k}$ and $\hat{b}_0$ prominently cascade perpendicular to $\hat{b}_0$, and such anisotropy increases with wavenumbers. The reduced power spectra of $\delta B_{\perp 1}$ follow Goldreich-Sridhar scalings: $\hat{P}(k_\perp) \propto k_\perp^{-5/3}$ and $\hat{P}(k_\parallel) \propto k_\parallel^{-2}$. In contrast, fluctuations within $\hat{k}\hat{b}_0$ plane show isotropic behaviors: perpendicular power distributions are approximately the same as parallel distributions. The reduced power spectra of fluctuations within the $\hat{k}\hat{b}_0$ plane follow the scalings: $\hat{P}(k_\perp) \propto k_\perp^{-3/2}$ and $\hat{P}(k_\parallel) \propto k_\parallel^{-3/2}$. Comparing frequency-wavevector spectra with theoretical dispersion relations of MHD modes, we find that $\delta B_{\perp 1}$ are probably associated with Alfvén modes. On the other hand, magnetic field fluctuations within the $\hat{k}\hat{b}_0$ plane more likely originate from fast modes based on their isotropic behaviors. The observations of anisotropy and scalings of different magnetic field components are consistent with the predictions of current compressible MHD theory. Moreover, for the Alfvén component, the ratio of cascading time to the wave period is found to be a factor of a few, consistent with critical balance in the strong turbulence regime. These results are valuable for further studies of energy compositions of plasma turbulence and their effects on energetic particle transports.

[1] Deutsches Elektronen Synchrotron DESY, Platanenallee 6, 15738, Zeuthen, Germany
[2] Institut für Physik und Astronomie, Universität Potsdam, 14476, Potsdam, Germany
[3] Department of Earth, Planetary, and Space Sciences, University of California, 90024, Los Angeles, California, USA
[4] LESIA, Observatoire de Paris, Université PSL, CNRS, Sorbonne Université, Université de Paris, 5 place Jules Janssen, 92195, Meudon, France
[5] Centre for Shandong Key Laboratory of Optical Astronomy and Solar-Terrestrial Environment, Institute of Space Sciences, 264209，Shandong University, Weihai, China
[6] Corresponding author huirong.yan@desy.de; terryliuzixu@ucla.edu

# 1. Introduction

Plasma turbulence is typically characterized by a broadband spectrum of perturbations, transmitting energy across a wide range of spatial and temporal scales (e.g., Bruno & Carbone 2013; Verscharen et al. 2019). Plasma turbulence plays a crucial role in solar corona, solar wind, fusion devices, and interstellar medium (e.g., Bruno & Carbone 2013; Yan & Lazarian 2008). The large-scale behaviors of plasma turbulence, which have been successfully described using magnetohydrodynamic (MHD) models, are of particular astrophysical interest (e.g., Kraichnan 1965; Goldreich & Sridhar 1995). The solar wind is easily accessed for in situ measurements of fields and particles, providing a unique laboratory for studying the physics of turbulent plasma observationally (e.g., Tu & Marsch 1995; Verscharen et al., 2019). Spacecraft observations and associated modeling have advanced our understanding of the solar wind in the last decades. However, turbulent properties and the three-dimensional structure of fluctuations remain unclear due to the limited number of sampling points and measurement difficulties. Thus, a study of the three-dimensional energy spectrum of the magnetic field is essential for understanding the dynamics of solar wind turbulence and their effects on energetic particle transports (e.g., Yan & Lazarian 2002; Yan & Lazarian 2004).

Solar wind fluctuations are anisotropic due to the presence of the local interplanetary magnetic field, which has been suggested in various studies (e.g., Matthaeus et al. 1990; Oughton et al. 2015). Satellite observations and simulations have shown the variance, power, and spectral index anisotropy in magnetic field components parallel and perpendicular to the field (e.g., Cho & Lazarian 2009; Oughton et al. 2015). First, solar wind fluctuations perpendicular to the background magnetic field ($B_0$) are typically more significant than parallel components, consistent with the dominance of incompressible Alfvén modes in the solar wind (e.g., Bruno & Carbone, 2013). Second, turbulence energy is predominantly transmitted perpendicular to $B_0$, based on the spatial correlation functions measured by single and multiple spacecraft (e.g., Matthaeus et al. 1990; He et al. 2011). Third, although solar wind fluctuations are often interpreted as a superposition of fluctuations with quasi-two-dimensional turbulence and a minority slab component, the perpendicular fluctuations are non-axisymmetric with respect to $B_0$, and preferentially in the direction perpendicular to $B_0$ and the radial direction (e.g., Bruno & Carbone 2013).

Theoretical progress has been achieved in understanding the anisotropic behaviors. Goldreich & Sridhar (1995) predicted that a scale-dependent anisotropy is present in incompressible strong MHD turbulence. The energy spectra of perpendicular and parallel components are $E(k_\perp) \propto k_\perp^{-5/3}$ and $E(k_\parallel) \propto k_\parallel^{-2}$, respectively. $k_\perp$ and $k_\parallel$ are wavenumbers perpendicular and parallel to $B_0$, respectively. The smaller turbulent eddies are more elongated along the local mean magnetic field (e.g., Cho & Lazarian 2009; Makwana & Yan 2020). A mechanism called three-wave resonant interaction also seems to be responsible for the anisotropy of magnetic field fluctuations (e.g., Shebalin et al. 1983; Cho & Lazarian 2002). Furthermore, according to the compressible MHD theory, plasma turbulence can be decomposed into three eigenmodes (Alfvén, slow, and fast modes) in a stationary, homogeneous, isothermal plasma with a uniform background magnetic field (e.g., Makwana & Yan 2020; Zhao et al.

2021). Using the term 'mode' in this study, we refer to the carriers of turbulent fluctuations, which are not dependent on the propagation properties of classical linear waves. In Fourier space, incompressible fluctuations whose displacement vectors are perpendicular to the $\hat{k}\hat{b}_0$ plane are defined as Alfvén modes, whereas fluctuations whose displacement vectors are within the $\hat{k}\hat{b}_0$ plane are defined as magnetosonic modes (slow and fast modes). The mode compositions of the turbulence can profoundly affect turbulence anisotropy (e.g., Yan & Lazarian 2004). Two-order structure functions show that the cascade of Alfvén and slow modes is anisotropic, preferentially in the direction perpendicular to the local magnetic field rather than the parallel direction, whereas fast modes tend to show isotropic cascade (e.g., Cho & Lazarian 2003; Makwana & Yan 2020). Moreover, both Alfvén and slow modes follow the Goldreich–Sridhar scalings, whereas fast modes follow isotropic scalings (Goldreich & Sridhar 1995; Cho & Lazarian 2003; Makwana & Yan 2020). Direct evidence from solar wind observations for the cascade of each mode is still lacking.

To investigate the anisotropy and scalings of solar wind turbulence with respect to the magnetic field at MHD scales at 1 au, we calculate three-dimensional power spectra in wavevector space using the Magnetospheric Multiscale (MMS) spacecraft (Burch et al. 2016). Narita et al. (2010) have tried to obtain three-dimensional energy distributions of magnetic field fluctuations using the wave telescope technique in an ordinary mean-field-aligned system. Compared with previous studies, this study organizes the three-dimensional magnetic power distributions in a new coordinate determined by wavevectors ($\hat{k}$) and background magnetic field ($\hat{b}_0$) in Fourier space. These measurements allow an examination of magnetic field fluctuations in terms of mode compositions. The organization of this paper is as follows. Section 2 describes data sets, analysis methods, and selection criteria. Section 3 offers observations. In Sections 4 and 5, we discuss and summarize our results.

## 2. Data and Methodology

### 2.1. Data

The study utilizes the magnetic field data from the fluxgate magnetometer (Russell et al. 2016) and the spectrograms of ion differential energy fluxes from the fast plasma investigation instrument (FPI; Pollock et al. 2016) onboard MMS. The resolutions of the survey-mode magnetic field measured by MMS are 8 samples/s and/or 16 samples/s. To provide sufficient time resolution for the timing analysis, we interpolate the magnetic field to a uniform time resolution of 64 samples/s. Due to FPI limitations in measuring plasma moments in the solar wind, proton parameters from the Operating Missions as a Node on the Internet (OMNI) are used to calculate the proton plasma $\beta_p$ (defined as $\beta_p \equiv \frac{P_p}{P_{mag}}$; $P_p$ = proton plasma pressure; $P_{mag}$ = magnetic pressure), proton gyro-frequency $f_{ci}$, proton gyro-radius $r_{ci}$, and proton inertial length $d_i$.

### 2.2. Analysis method

The magnetic field observed by four MMS spacecraft consists of the background and fluctuating magnetic field, i.e., $\boldsymbol{B} = \boldsymbol{B}_0 + \delta\boldsymbol{B}$. The background magnetic field is obtained by averaging the magnetic field within the defined time window, $\boldsymbol{B}_0 = \langle \boldsymbol{B} \rangle$. We calculate three-dimensional power spectra of magnetic field fluctuations with the following steps.

First, the time series of the fluctuating magnetic field is transformed into Fourier space by the Morlet-wavelet transforms (Grinsted et al. 2004). We obtain wavelet coefficients of three components of the fluctuating magnetic field, i.e., $W_{B_{X,GSE}}(t, f_{sc})$, $W_{B_{Y,GSE}}(t, f_{sc})$, and $W_{B_{Z,GSE}}(t, f_{sc})$ at each time $t$ and spacecraft-frame frequency ($f_{sc}$), where the subscript $GSE$ represents the geocentric-solar-ecliptic coordinates. We utilize the intervals with twice the length of the studied period to eliminate the edge effect due to finite-length time series and cut off the affected periods.

Second, we calculate unit wavevectors $\widehat{\boldsymbol{k}}(t, f_{sc}) = \frac{\boldsymbol{k}}{|\boldsymbol{k}|}$ using the singular value decomposition (SVD) of the magnetic spectral matrix (Santolík et al. 2003; Zhao et al. 2021). The SVD technique provides a mathematical method to solve the linearized Gauss's law for magnetism ($\boldsymbol{k} \cdot \delta \boldsymbol{B} = 0$) that states a divergence-free constraint of magnetic field vectors. The complex matrices of $\delta \boldsymbol{B}$ in this study are expressed as the wavelet coefficients. The wavevectors $\widehat{\boldsymbol{k}}$ are calculated by the SVD method with 32 s resolution since we find that the wavevectors of low-frequency fluctuations are relatively stationary with varying time resolution.

Third, since the relative separations of MMS spacecraft are much smaller than the half-wavelength at MHD scales, the properties and propagations of fluctuations from simultaneous measurements by four MMS spacecraft are approximately similar. The wavevectors and background magnetic field are averaged over four spacecraft: $\boldsymbol{k} = \frac{1}{4}(\widehat{\boldsymbol{k}}_{M1} + \widehat{\boldsymbol{k}}_{M2} + \widehat{\boldsymbol{k}}_{M3} + \widehat{\boldsymbol{k}}_{M4})$ and $\boldsymbol{B}_0 = \frac{1}{4}(\boldsymbol{B}_{0,M1} + \boldsymbol{B}_{0,M2} + \boldsymbol{B}_{0,M3} + \boldsymbol{B}_{0,M4})$, where $M1$, $M2$, $M3$, and $M4$ denote the four MMS spacecraft. We build a new coordinate in Fourier space using the unit vectors of the wavevectors and background magnetic field ($\widehat{\boldsymbol{k}} = \frac{\boldsymbol{k}}{|\boldsymbol{k}|}$ and $\widehat{\boldsymbol{b}}_0 = \frac{\boldsymbol{B}_0}{|\boldsymbol{B}_0|}$) at each time $t$ and $f_{sc}$, where the basis vectors of coordinate axes $\boldsymbol{e}_{\parallel}$, $\boldsymbol{e}_{\perp 1}$, and $\boldsymbol{e}_{\perp 2}$ are in $\widehat{\boldsymbol{b}}_0$, $\widehat{\boldsymbol{k}} \times \widehat{\boldsymbol{b}}_0$, and $\widehat{\boldsymbol{b}}_0 \times (\widehat{\boldsymbol{k}} \times \widehat{\boldsymbol{b}}_0)$ directions, respectively. Then, wavelet coefficients are transformed into the new coordinate: $W_{B_{\parallel}}(t, f_{sc})$, $W_{B_{\perp 1}}(t, f_{sc})$, and $W_{B_{\perp 2}}(t, f_{sc})$.

Fourth, four MMS spacecraft provide six cross correlations for the magnetic field, i.e., $W_{B_l}^{12} = \langle W_{B_l,M1} W_{B_l,M2}^* \rangle$, $W_{B_l}^{13} = \langle W_{B_l,M1} W_{B_l,M3}^* \rangle$, $W_{B_l}^{14} = \langle W_{B_l,M1} W_{B_l,M4}^* \rangle$, $W_{B_l}^{23} = \langle W_{B_l,M2} W_{B_l,M3}^* \rangle$, $W_{B_l}^{24} = \langle W_{B_l,M2} W_{B_l,M4}^* \rangle$, and $W_{B_l}^{34} = \langle W_{B_l,M3} W_{B_l,M4}^* \rangle$ (Grinsted et al. 2004), where the angular brackets denote time averaging over 32 s for each $f_{sc}$, and the subscript $l$ represents $\boldsymbol{e}_{\parallel}$, $\boldsymbol{e}_{\perp 1}$, and $\boldsymbol{e}_{\perp 2}$, respectively. We calculate wavevectors $\boldsymbol{k}_l(t, f_{sc})$ employing the multi-spacecraft timing analysis based on phase differences between the fluctuating magnetic field (with a time resolution of 64 samples/s) as measured by four spacecraft (Pinçon & Glassmeier 2008).

Fifth, wavelet power spectra of the magnetic field averaged over four spacecraft at each time $t$ and $f_{sc}$ in component $l$ are given by

$$P_{B_l}(t, f_{sc}) = \frac{1}{4}\left(W_{B_l,M1}W_{B_l,M1}^* + W_{B_l,M2}W_{B_l,M2}^* + W_{B_l,M3}W_{B_l,M3}^* + W_{B_l,M4}W_{B_l,M4}^*\right) \qquad (1).$$

Combining $P_{B_l}(t, f_{sc})$ and $\boldsymbol{k}_l(t, f_{sc})$, we obtain frequency-wavenumber power spectra of magnetic field fluctuations $P_{B_l}(\boldsymbol{k}_l, f_{sc})$ in the spacecraft frame.

Finally, magnetic power spectra are transformed into the rest frame of the solar wind by correcting the Doppler shift. The frequency in the rest frame of the solar wind can be obtained as $f_{rest} = f_{sc} - \frac{\mathbf{k} \cdot \mathbf{V}}{2\pi}$. Compared to the solar wind flow, four spacecraft are approximately stationary. Thus, $\mathbf{V}$ is roughly equivalent to the solar wind speed ($V_{sw}$). This study exploits the representation of absolute frequencies:

$$(f_{rest}, \mathbf{k}_l) = \begin{cases} (f_{rest}, \mathbf{k}_l), & \text{for } f_{rest} > 0 \\ (-f_{rest}, -\mathbf{k}_l), & \text{for } f_{rest} < 0 \end{cases}.$$

### 2.3. selection criteria

We search for events that satisfy the following criteria: (1) The spectrograms of differential energy fluxes show no evidence of high-energy reflected ions from the terrestrial bow shock, suggesting that fluctuations are in the free solar wind without the effects of the ion foreshock. (2) The magnetic field is devoid of strong gradients, discontinuities, and reversals, guaranteeing that plasma can be considered homogeneous. (3) The fluctuating magnetic field is smaller than the background magnetic field, and relative amplitudes of magnetic field $\delta B_{rms}/|\mathbf{B}_0| = \frac{\sqrt{\langle |\mathbf{B}(t) - \langle \mathbf{B}(t) \rangle|^2 \rangle}}{|\langle \mathbf{B}(t) \rangle|} < 1$. Under such a condition, the nonlinear term ($\delta B^2$) is much less than the linear term ($\mathbf{B}_0 \cdot \delta \mathbf{B}$), and thus fluctuations can be approximately considered as a superposition of MHD eigenmodes (Alfvén, fast, and slow modes). (4) Multi-spacecraft methods are sensitive to scales comparable to spacecraft separations and show limitations on much larger and smaller scales (e.g., Horbury et al. 2012). In this study, the spacecraft separations are roughly comparable to the proton inertial length ($d_i$). Therefore, given the applicability of MHD theory and measurement limitation, we only analyze fluctuations within $\frac{2}{t^*} < f_{rest} < f_{ci}$ and $k d_i < 0.2$, and set the magnetic power to zero out of this range. The parameter $t^*$ is the duration studied.

The only other criterion used is the requirement that the angle $\phi_{\hat{k}k_l}$ between $\hat{\mathbf{k}}$ (obtained by SVD method) and $\mathbf{k}_l$ (obtained by timing analysis) should be small. Typically, when analyzing solar wind data via single spacecraft, we assume $f_{sc} \propto k$ based on Taylor's hypothesis (Taylor 1938) for a given wave propagation angle $\theta_{kB_0}$. This approximation is considered reliable mainly because the velocity of the solar wind flow ($V_{sw}$) is much larger than the phase speeds of MHD waves. However, the approximation does not always hold even though $V_{sw}$ is much faster than the phase speed of fluctuations, e.g., when fluctuations have wavevectors at large angles from the solar wind flow. Therefore, this study utilizes two methods for accuracy to identify the propagation directions, i.e., the SVD method and multi-spacecraft timing analysis (described in Section 2.2). The latter method allows determining wavevectors independent of any spatiotemporal hypothesis. Our results show that data counts are primarily concentrated in $\phi_{\hat{k}k_l} < 30°$, whereas a small number of counts still exist in large-$\phi_{\hat{k}k_l}$ range. It indicates that not all fluctuations are aligned with the direction of minimum variance vectors of magnetic field fluctuations and satisfy $f_{sc} \propto k$ hypothesis (because the fluctuations are combinations of multiple modes with different dispersion relations). Therefore, we filter out fluctuations with a large $\phi_{\hat{k}k_l}$, which invalidates the SVD assumption. Such fluctuations are beyond the scope of the present paper and will be the topic of a separate publication. Considering that the fluctuating magnetic field ($\delta B_{\perp 1}$) out of the $\hat{k}\hat{b}_0$ plane dominates magnetic power (~80%), the propagation direction of $\delta B_{\perp 1}$ should be mainly aligned with $\hat{k}$, whereas fluctuations within the $\hat{k}\hat{b}_0$ plane accounting for a tiny proportion have little impact on the direction of $\hat{k}$. Therefore, this study sets a more stringent criterion $\phi_{\hat{k}k_l} < 10°$ for $\delta B_{\perp 1}$ fluctuations and a moderate criterion $\phi_{\hat{k}k_l} < 30°$ for fluctuations within the $\hat{k}\hat{b}_0$ plane.

This study presents three representative events in sub-Alfvénic solar wind turbulence at 1 AU, and their properties are listed in Table 1. These three events are also included in the appendix by Roberts et al. (2020). During these intervals, four spacecraft have a ~1-minute time shift from the nose of the terrestrial bow shock, suggesting approximately the same plasma environment observed by MMS and OMNI. Meanwhile, the qualities of MMS tetrahedral configuration are around 0.9, allowing for distinguishing spatial and temporal evolutions and investigating three-dimensional structures of fluctuations.

### 3. Observations

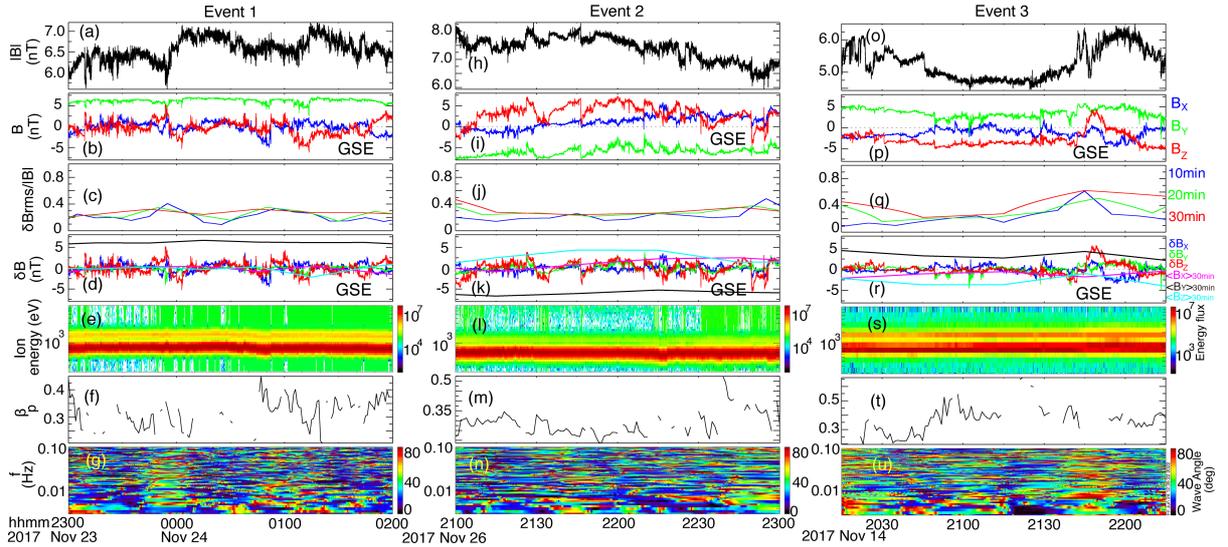

**Figure 1.** An overview of solar wind fluctuations from MMS1 and OMNI. (a–g) Event 1: during 23:00–02:00 UT on 2017 November 23–24. (h–n) Event 2: during 21:00–23:00 UT on 2017 November 26. (o–u) Event 3: during 20:15–22:15 UT 2017 November 14. The panels from top to bottom show the magnetic field magnitude $|B|$, magnetic field components in GSE, relative amplitudes of the magnetic field $\delta B_{rms}/|B_0|$, comparisons of fluctuating magnetic field and background magnetic field in GSE, the spectrogram of ion differential energy fluxes (in units of $keV\,cm^{-2}\,s^{-1}\,sr^{-1}\,keV^{-1}$), proton plasma $\beta_p$, wave propagation angles with respect to the background magnetic field.

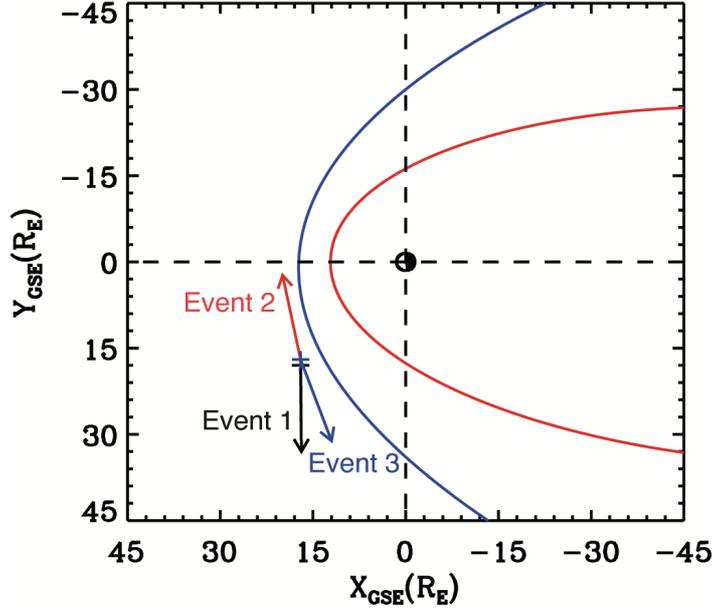

**Figure 2.** A schematic of MMS locations in GSE coordinates. Blue and red solid curves represent the empirical models of terrestrial bow shock (Peredo et al. 1995) and magnetopause (Roelof & Sibeck 1993), respectively. The black, red, and blue arrows represent the directions of the background magnetic field for Event 1 ([17,18,6] $R_E$), Event 2 ([17,17,6] $R_E$), and Event 3 ([17,17,6] $R_E$), respectively.

An overview of three representative events of solar wind fluctuations is shown in Figure 1. For all three events, the magnetic field and plasma parameters are stationary. Figures 1c, 1j, and 1q show that relative amplitudes of the magnetic field $\delta B_{rms}/|\boldsymbol{B}_0| = \sqrt{\langle |\boldsymbol{B}(t) - \langle \boldsymbol{B}(t)\rangle|^2\rangle}/|\langle\boldsymbol{B}(t)\rangle|$ are less than unity, where the angular brackets denote a time average over 10, 20, and 30 minutes, respectively. Figures 1d, 1k, and 1r compare the fluctuating magnetic field and background magnetic field $\langle \boldsymbol{B}\rangle_{30\,min}$ (average over 30 minutes). Given the small fluctuations with a strong background magnetic field, it is valid to linearize MHD equations, ignoring the second- and higher-order contributions. Thus, the fluctuations can be approximately considered as a superposition of MHD eigenmodes (Alfvén, fast, and slow modes). Figure 2 shows MMS locations and the directions of the background magnetic field in GSE coordinates for Event 1, Event 2, and Event 3, respectively. As shown in Figure 2, theoretically, the mean magnetic field is either not connected to terrestrial bow shock or nearly tangential to it. Indeed, the spectrograms of ion differential energy fluxes show no evidence of high-energy reflected ions (Figures 1e, 1l, and 1s). Thus, these intervals are free from ion foreshock contaminations. Figures 1f, 1m, and 1t show that average $\beta_p$ is around 0.3, where proton plasma $\beta_p$ is calculated by OMNI proton parameters and MMS magnetic field. In Figures 1g, 1n, and 1u, wave propagation angles with respect to the background magnetic field $\langle \boldsymbol{B}\rangle_{30\,min}$ cover $0°$-$90°$, allowing us to calculate the magnetic power distributions in wavevector space more reliable.

The solar wind fluctuations are closely related to the local background magnetic field. This study explores the variation of the magnetic power distributions with the local background magnetic field by adjusting the length of time windows. We split the time intervals of these three events into several moving time windows with a step size of 5 minutes and a length of 10 minutes, 20 minutes, and 30 minutes, respectively. This study refers to them as 10 minute, 20 minute, and 30 minute data sets. We calculate the background magnetic field ($B_0$) by averaging the magnetic field in each time window. In this way, $B_0$ in each time window is constant in time and along the same direction in both real and Fourier space, suggesting that the new coordinate determined by $B_0$ is independent of the space transformation.

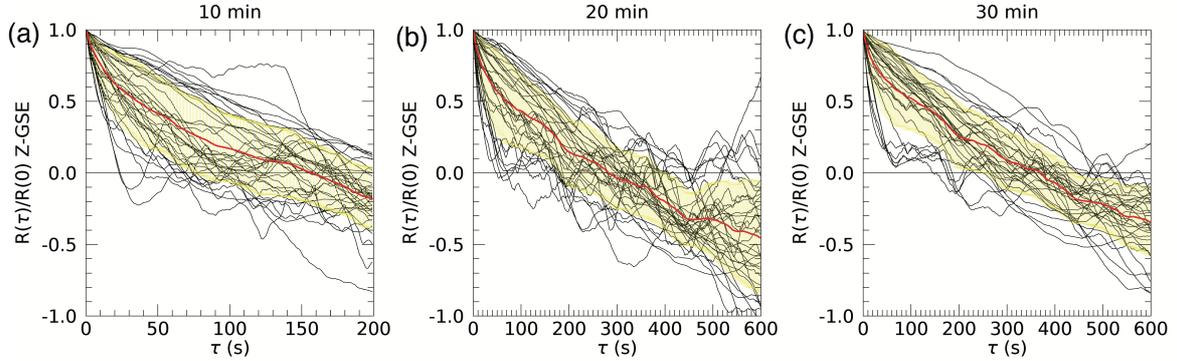

**Figure 3.** Normalized correlation function $R(\tau)/R(0)$ vs. timescale $\tau$ for magnetic field $Z_{GSE}$ component with window lengths of (a) 10 minutes, (b) 20 minutes, and (c) 30 minutes. The black curves represent $R(\tau)/R(0)$ from different moving time windows. The red curves represent average values ($s$) of $R(\tau)/R(0)$ over the corresponding time windows. The yellow shaded regions represent $[s - \sigma, s + \sigma]$, where $\sigma$ is the standard deviation. The horizontal lines mark mark $R(\tau)/R(0) = 0$.

To investigate the reliability of employing observational three-dimensional power distributions to describe the structure of solar wind turbulence, we calculate the normalized two-point correlation function ($R(\tau)/R(0)$), where $R(\tau)$ is defined as $\langle \delta B(t) \delta B(t + \tau) \rangle$ and $\tau$ is the timescale. The three events show similar results; thus, we take Event 1 as an example. Figures 3a–3c show $R(\tau)/R(0)$ for the magnetic field $Z_{GSE}$ component obtained from data sets with lengths of 10, 20, and 30 minutes, respectively. The black curves represent $R(\tau)/R(0)$ calculated from magnetic field data in moving time windows with a step size of 5 minutes. The profiles of $R(\tau)/R(0)$ obtained from different windows mainly appear within the yellow shaded region $[s - \sigma, s + \sigma]$, where $s$ and $\sigma$ represent the average value and standard deviation of $R(\tau)/R(0)$ over the corresponding time windows, respectively. The result indicates that the correlation functions do not depend on the starting time of moving time windows; thus, fluctuations are in a homogeneous plasma. Moreover, when well-behaved turbulence becomes uncorrelated ($R(\tau)/R(0) \to 0$), the average values of $\tau$ are around 160 s, 280 s, and 370 s, respectively. Therefore, the measured correlation time $T_c$, estimated by $T_c = \int_0^{R(\tau) \to 0} R(\tau)/R(0) d\tau$, is much less than the window length. As a result, it is reliable to assume that turbulent magnetic field fluctuations are stationary and homogeneous (Matthaeus & Goldstein 1982).

We follow the approaches described in Section 2.2 to calculate three-dimensional frequency-wavenumber magnetic power spectra in the spacecraft frame. Then we transform the magnetic power spectra into the rest frame of the solar wind by correcting the Doppler shift. To obtain $k_\perp - k_\parallel$ magnetic power spectra, we construct a set of 100×100 bins, where $k_\parallel$ represents the wavenumber parallel to the background magnetic field ($\boldsymbol{B_0}$), and $k_\perp = \sqrt{k_{11}^2 + k_{12}^2}$ represents the wavenumber perpendicular to $\boldsymbol{B_0}$. Each bin subtends approximately the same perpendicular and parallel wavenumber. We sum all magnetic power in each bin at all frequencies and times. To cover all MHD-scale wavenumbers, we set the maximum wavenumber as $k_{max} = k_{\perp,max} = k_{\parallel,max} = 1.1 \times \frac{0.2}{d_i}$, and the step length of each bin is $dk = \frac{k_{max}}{100}$.

### 3.1. The Results of $\delta B_{\perp 1}$ Fluctuations

**Event 1**: 2017-11-(23-24)/23:00-02:00 UT

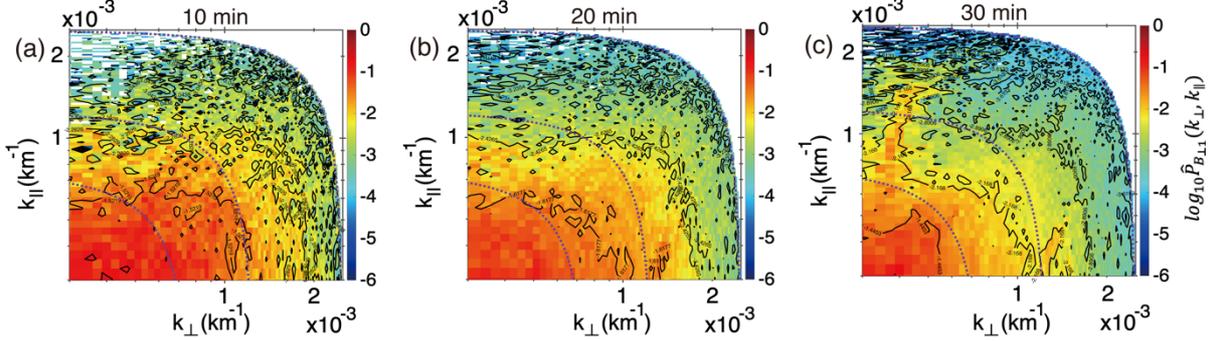

**Event 2**: 2017-11-26/21:00-23:00 UT

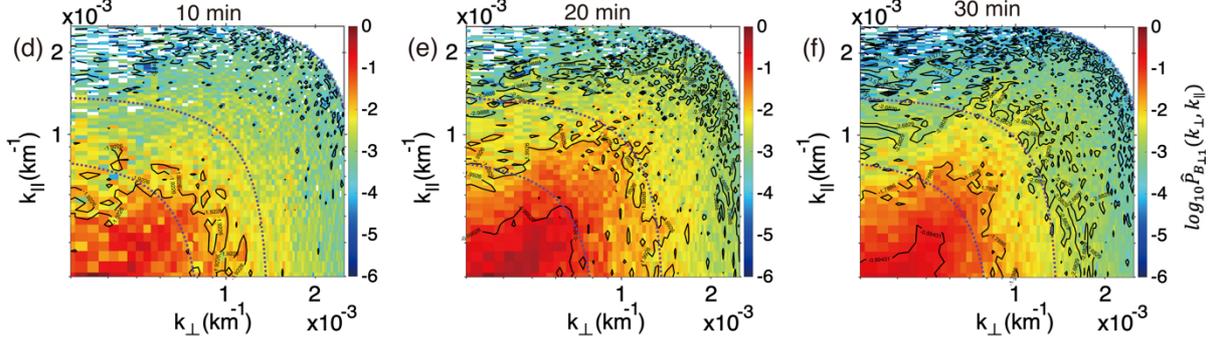

**Event 3**: 2017-11-14/20:15-22:15 UT

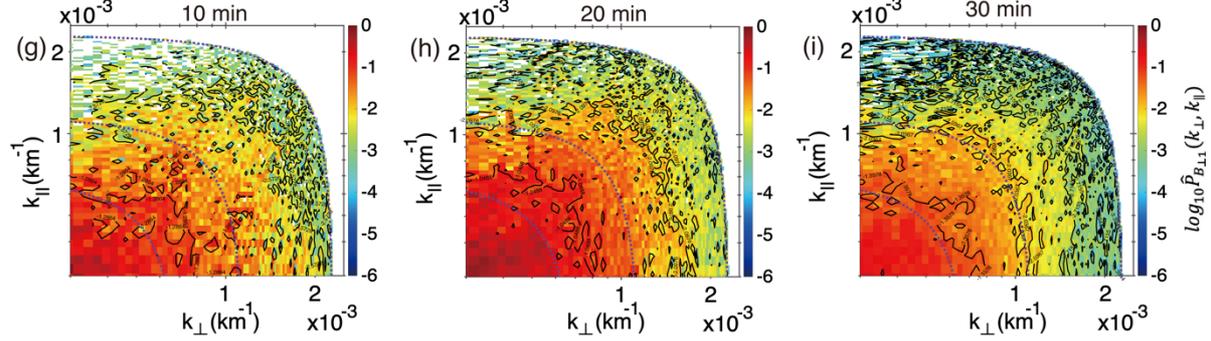

**Figure 4.** $k_\perp - k_\parallel$ wavelet power spectra of magnetic field fluctuations ($\delta B_{\perp 1}$) out of the $\hat{k}\hat{b}_0$ plane using data sets with lengths of 10 minutes (left), 20 minutes (middle), and 30 minutes (right). The magnetic power $\hat{P}_{B_{\perp 1}}(k_\perp, k_\parallel) = P_{B_{\perp 1}}(k_\perp, k_\parallel)/P_{B_{\perp 1},max}$ are normalized by the maximum power in all bins. (a–c) Event 1: during 23:00–02:00 UT on 2017 November 23–24. (d–f) Event 2: during 21:00–23:00 UT on 2017 November 26. (g–i) Event 3: during 20:15–22:15 UT 2017 November

November 14. The purple dashed curves mark $k = \sqrt{k_\parallel^2 + k_\perp^2} = 0.06/d_i$, $0.1/d_i$, and $0.2/d_i$.

Figure 4 shows $k_\perp - k_\parallel$ wavelet power spectra of magnetic field fluctuations ($\delta B_{\perp 1}$) out of the $\hat{k}\hat{b}_0$ plane using data sets with lengths of 10 minutes (left), 20 minutes (middle), and 30 minutes (right). The magnetic power $\hat{P}_{B_{\perp 1}}(k_\perp, k_\parallel) = P_{B_{\perp 1}}(k_\perp, k_\parallel)/P_{B_{\perp 1}, max}$ is normalized by the maximum power in all bins. From top to bottom, Figures 4a-c, d-f, and g-i display magnetic power spectra of Event 1, Event 2, and Event 3, respectively. For all events, the magnetic power spectra are prominently distributed along the $k_\perp$ axis, indicating a faster cascade in the perpendicular direction. Moreover, we observe an apparent scale-dependent anisotropy: magnetic power spectra are more stretched along the $k_\perp$ axis with increasing wavenumbers. Besides, magnetic power spectra show more isotropic behaviors as the window length increases.

To quantitatively analyze these properties of magnetic power spectra, we obtain the one-dimensional reduced magnetic power spectra of $\delta B_{\perp 1}$ as

$$\hat{P}_{B_{\perp 1}}(k_\perp) = \sum_{k_\parallel = 0}^{k_\parallel \to \infty} \hat{P}_{B_{\perp 1}}(k_\perp, k_\parallel) \sim \sum_{k_\parallel = k_{\parallel,min}}^{k_\parallel = k_{max}} \hat{P}_{B_{\perp 1}}(k_\perp, k_\parallel), \tag{2}$$

$$\hat{P}_{B_{\perp 1}}(k_\parallel) = \sum_{k_\perp = 0}^{k_\perp \to \infty} \hat{P}_{B_{\perp 1}}(k_\perp, k_\parallel) \sim \sum_{k_\perp = k_{\perp,min}}^{k_\perp = k_{max}} \hat{P}_{B_{\perp 1}}(k_\perp, k_\parallel), \tag{3}$$

The parallel wavenumber can be expressed as $k_\parallel \propto k_\perp^{\frac{2}{3}}/l_0^{\frac{1}{3}}$, where $l_0$ is the injection scale (approximately equivalent to the correlation length in this study) (Yan & Lazarian 2008). Here, we take Event 1 as an example to estimate the minimum wavenumber. Given that $\frac{\delta V}{V_A}$ can be roughly expressed as $\frac{\delta B}{B_0}$ for incompressible fluctuations, velocity fluctuations $\delta V$ are around 15 km/s, where Alfvén velocity $V_A \sim 50 \, km/s$ and $\frac{\delta B}{B_0} \sim \langle \frac{\delta B_{rms}}{|B|} \rangle_{30min} \sim 0.3$. The correlation time $T_c$ stabilizes at around $400 \, s$, even though using longer time windows (larger than 30 minutes). Thus, the correlation length is around 6000 km. If we assume the minimum perpendicular wavenumber $k_{\perp,min} \sim 0.1 \times k_{\parallel,max} \sim 2.5 \times 10^{-4} \, km^{-1}$, the minimum parallel wavenumber is $k_{\parallel,min} \sim 2.2 \times 10^{-4} \, km^{-1} \sim k_{\perp,min}$. The calculations can be confirmed by the relationship between $k_\perp$ and $k_\parallel$ at the small wavenumbers in Figure 10d.

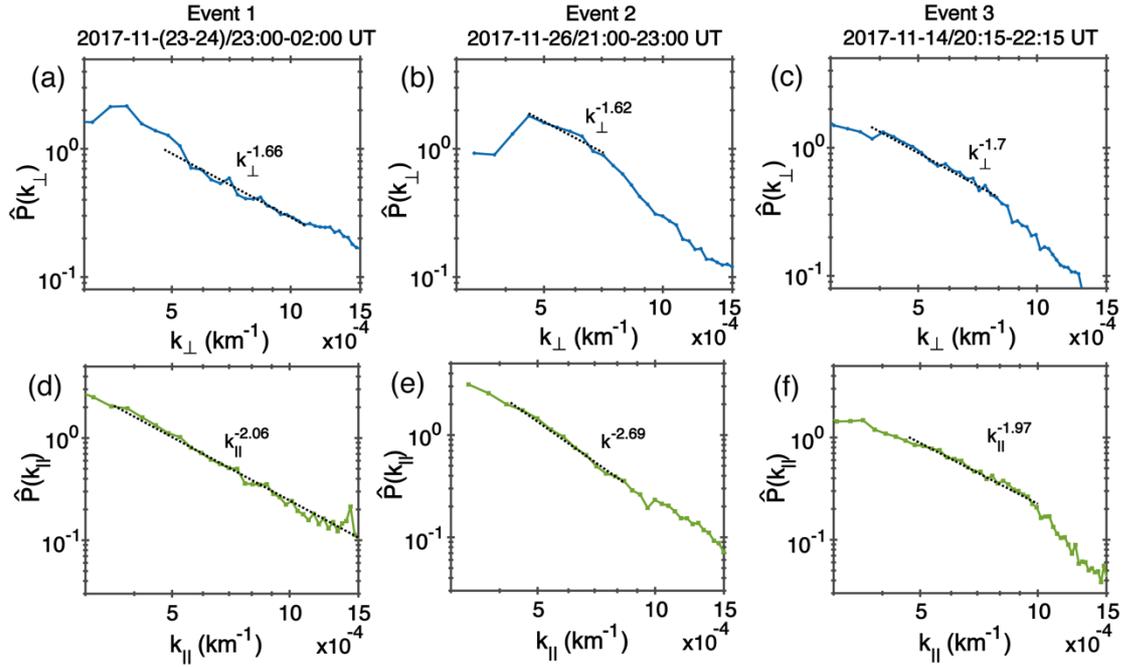

**Figure 5.** The reduced power spectra of $\delta B_{\perp 1}$ fluctuations using 30 minute data sets. Blue curve $\hat{P}(k_\perp)$: normalized perpendicular wavenumber spectrum of $\delta B_{\perp 1}$ fluctuations; green curve $\hat{P}(k_\parallel)$: normalized parallel wavenumber spectrum of $\delta B_{\perp 1}$ fluctuations; black dashed line: power-law fits. (a, d) Event 1: during 23:00-02:00 UT on 2017 November 23-24. (b, e) Event 2: during 21:00-23:00 UT on 2017 November 26. (c, f) Event 3: during 20:15-22:15 UT 2017 November 14.

It is challenging to quantitatively present magnetic power distributions in wavevector space through limited measurements. However, the longer the intervals we choose, the closer the observed power spectra are to actual distributions. Thus, we present reduced power spectra of $\delta B_{\perp 1}$ fluctuations using 30 minute data sets, although shorter data sets show similar behaviors. Figures 5a-5c show that the normalized perpendicular wavenumber spectra of $\delta B_{\perp 1}$ fluctuations roughly follow a Kolmogorov spectrum: $\hat{P}(k_\perp) \propto k_\perp^{-5/3}$. Figures 5d-5f show that the normalized parallel wavenumber spectra of $\delta B_{\perp 1}$ fluctuations roughly follow $\hat{P}(k_\parallel) \propto k_\parallel^{-2}$. These scalings are consistent with the Goldreich & Sridhar (1995) theory. The reduced power spectra reveal more clearly the tendency that fluctuations prominently cascade perpendicular to the background magnetic field. For event 1 (3 hr measurements), magnetic power distributions almost follow the Goldreich & Sridhar (1995) scalings from $k_\parallel \sim [4 \times 10^{-4}, 1.5 \times 10^{-3}]\ km^{-1}$ (corresponding wavenumber relations are shown in Figure 10d). However, for Events 2 and 3 (2 hr measurements), magnetic power distributions satisfy the scale-dependence scaling at partial wavenumbers. Their power-law fits are easily affected by the wavenumber ranges, likely because of the incomplete results from the limited-time series.

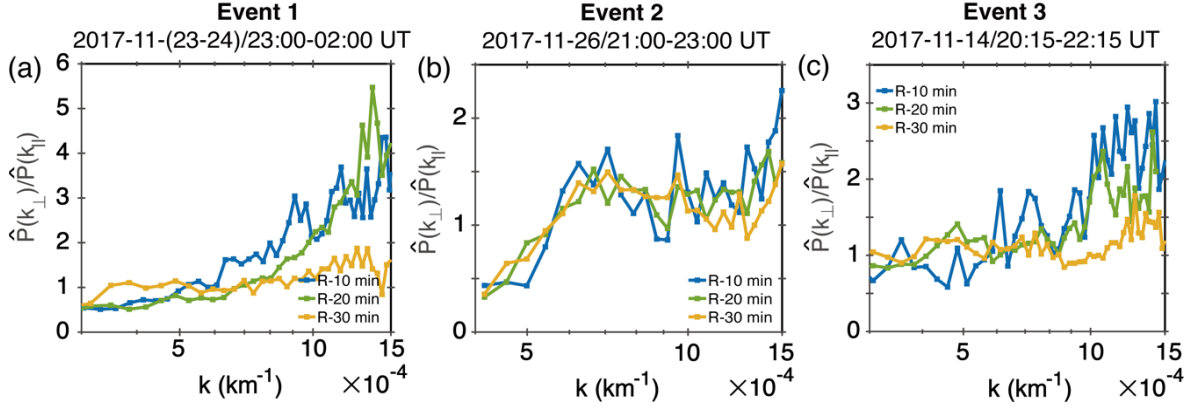

**Figure 6.** The ratio $\hat{P}(k_\perp)/\hat{P}(k_\parallel)$ of $\delta B_{\perp 1}$ fluctuations. $\hat{P}(k_\perp)$ represents the normalized perpendicular wavenumber spectrum of $\delta B_{\perp 1}$, and $\hat{P}(k_\parallel)$ represents the normalized parallel wavenumber spectrum of $\delta B_{\perp 1}$. (a) Event 1: during 23:00–02:00 UT on 2017 November 23–24. (b) Event 2: during 21:00–23:00 UT on 2017 November 26. (c) Event 3: during 20:15–22:15 UT 2017 November 14.

To investigate the anisotropy of magnetic power spectra of $\delta B_{\perp 1}$ fluctuations, we show the ratio $\hat{P}(k_\perp)/\hat{P}(k_\parallel)$ in Figure 6. First, the ratios $\hat{P}(k_\perp)/\hat{P}(k_\parallel)$ are much larger than one at most wavenumbers, indicating a faster cascade in the perpendicular direction. Second, almost all data sets show a similar tendency that the ratio $\hat{P}(k_\perp)/\hat{P}(k_\parallel)$ increases with the wavenumbers, especially for $k > 5 \times 10^{-4}\,km^{-1}$, suggesting that the anisotropy of magnetic power spectra increases with the wavenumber. This result is consistent with simulation results: the smaller eddies are more stretched along the background magnetic field (Makwana & Yan 2020). Third, the ratios $\hat{P}(k_\perp)/\hat{P}(k_\parallel)$ obtained by 10 minutes (blue) and 20 minutes (green) data sets are larger than those obtained by 30 minute data sets (yellow) at most wavenumbers, especially for $k > 5 \times 10^{-4}\,km^{-1}$. It means that the anisotropy decreases when we use the longer time windows to calculate the background magnetic field. Therefore, these observations provide evidence that solar wind fluctuations are more likely aligned with the *local* magnetic field based on the size of the fluctuations rather than a global magnetic field.

### 3.2. The Results of Fluctuations within the $\hat{k}\hat{b}_0$ Plane

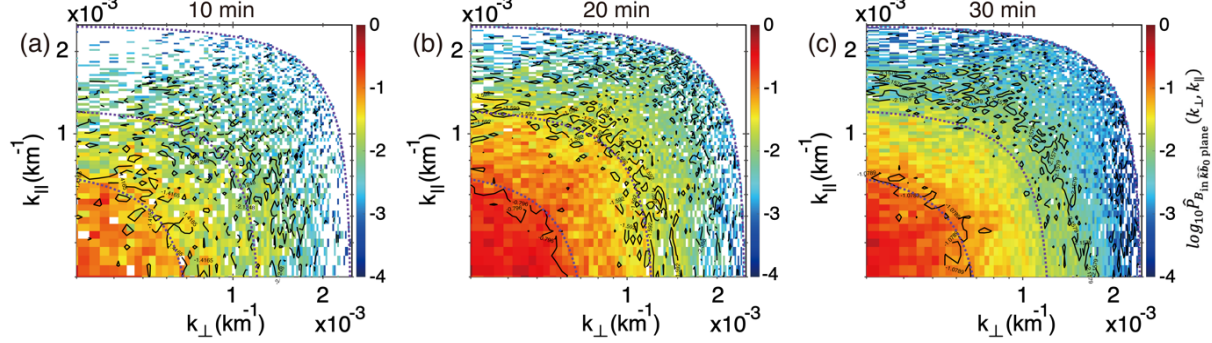

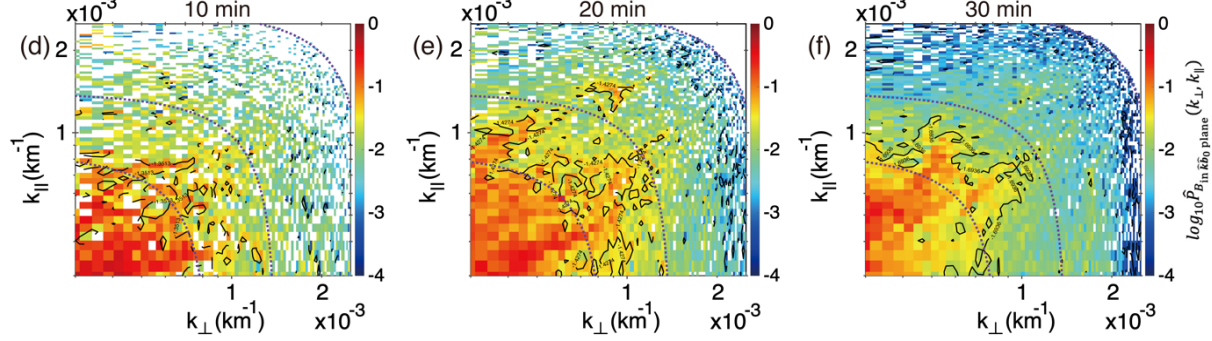

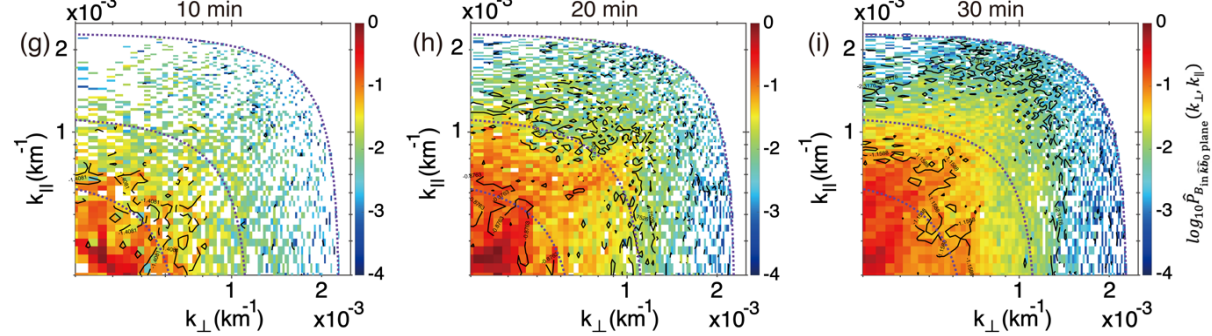

**Figure 7.** $k_\perp - k_\parallel$ wavelet power spectra of magnetic field fluctuations within the $\hat{k}\hat{b}_0$ plane using data sets with lengths of 10 minutes (left), 20 minutes (middle), and 30 minutes (right). The magnetic power $\hat{P}_{B_{\text{in }\hat{k}\hat{b}_0\text{ plane}}}(k_\perp, k_\parallel) = (P_{B_\parallel}(k_\perp, k_\parallel) + P_{B_{\perp 2}}(k_\perp, k_\parallel))/(P_{B_\parallel} + P_{B_{\perp 2}})_{max}$ is normalized by the maximum power in all bins. (a-c) Event 1: during 23:00-02:00 UT on 2017 November 23-24. (d-f) Event 2: during 21:00-23:00 UT on 2017 November 26. (g-i) Event 3: during 20:15-22:15 UT 2017 November 14. The purple dashed curves mark $k = 0.06/d_i$, $0.1/d_i$, and $0.2/d_i$.

The magnetic field fluctuations within the $\hat{k}\hat{b}_0$ plane are composed of $\delta B_\parallel$ and $\delta B_{\perp 2}$ components, where $\delta B_\parallel$ represents fluctuations parallel to $\boldsymbol{B_0}$, and $\delta B_{\perp 2}$ represents fluctuations perpendicular to $\boldsymbol{B_0}$ and within the $\hat{k}\hat{b}_0$ plane. Based on the ideal MHD theory, these magnetic field fluctuations are provided by compressible magnetosonic modes. Figure 7 presents the sum of $k_\perp - k_\parallel$ wavelet power spectra of $\delta B_\parallel$ and $\delta B_{\perp 2}$ fluctuations, which are normalized by the maximum power in all bins ($\hat{P}_{B_{\text{in }\hat{k}\hat{b}_0\text{ plane}}}(k_\perp, k_\parallel) = (P_{B_\parallel}(k_\perp, k_\parallel) + P_{B_{\perp 2}}(k_\perp, k_\parallel))/(P_{B_\parallel} + P_{B_{\perp 2}})_{max}$). Since these sub-Alfvénic fluctuations within the $\hat{k}\hat{b}_0$ plane only occupy a tiny part of the total magnetic power (~20%), magnetic power cannot cover all wavenumbers. Thus, many vacant bins are present in Figure 7. Nevertheless, magnetic power spectra within the $\hat{k}\hat{b}_0$ plane still show explicit isotropic behaviors: the perpendicular magnetic power distributions are comparable to those in the parallel direction.

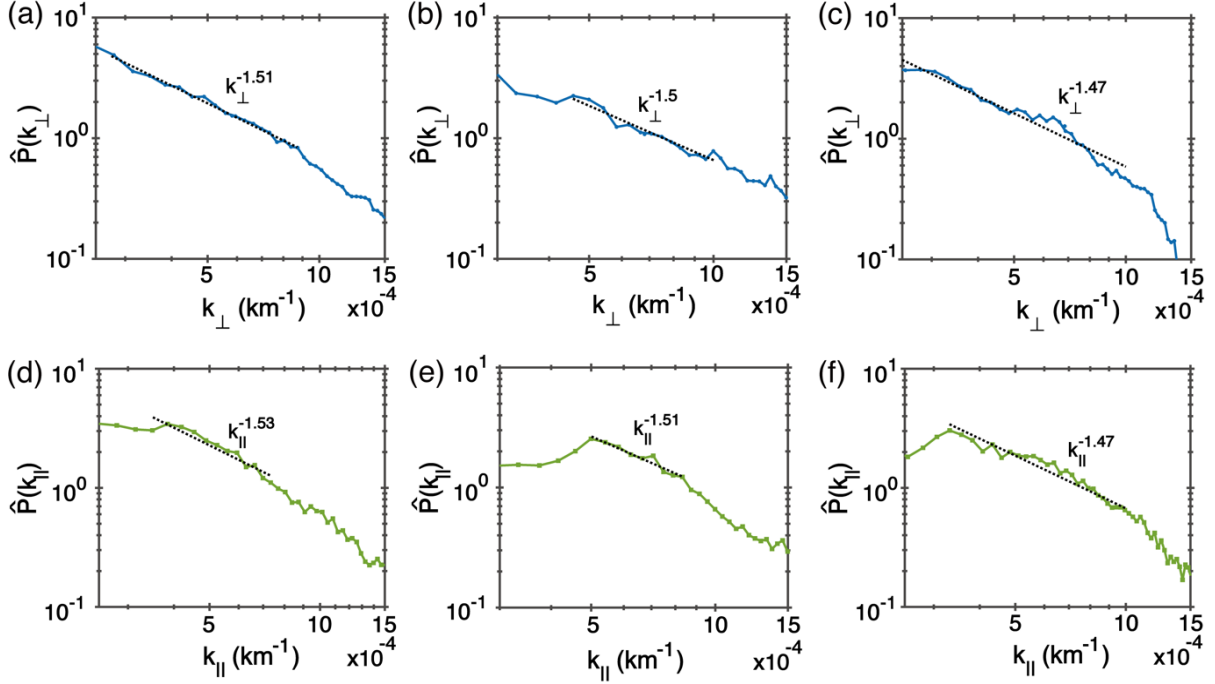

**Figure 8.** The reduced power spectra of magnetic field fluctuations within $\hat{k}\hat{b}_0$ plane using 30 minute data sets. Blue curve $\hat{P}(k_\perp)$: normalized perpendicular wavenumber spectrum; green curve $\hat{P}(k_\parallel)$: normalized parallel wavenumber spectrum; black dashed lines: power-law fits. (a,d) Event 1: during 23:00-02:00 UT on 2017 November 23-24. (b,e) Event 2: during 21:00-23:00 UT on 2017 November 26. (c,f) Event 3: during 20:15-22:15 UT 2017 November 14.

We calculate one-dimensional reduced power spectra of magnetic field fluctuations within the $\hat{k}\hat{b}_0$ plane using Equations (2) and (3) for in-plane components with 30 minute data sets. Although the reliability of quantitative analysis is questionable, the normalized perpendicular wavenumber spectra $\hat{P}(k_\perp)$ are roughly comparable to the parallel wavenumber spectra $\hat{P}(k_\parallel)$ (Figure 8), and the ratios $\hat{P}(k_\perp)/\hat{P}(k_\parallel)$ are around one (Figure 9). Therefore, the isotropic behaviors are independent of the wavenumbers. Moreover, the reduced power spectra follow a similar scaling $\hat{P}(k_\perp) \propto k_\perp^{-3/2}$ and $\hat{P}(k_\parallel) \propto k_\parallel^{-3/2}$ in Figure 8. The isotropic scalings are consistent with fast-mode scalings (Cho & Lazarian 2003; Makwana & Yan 2020).

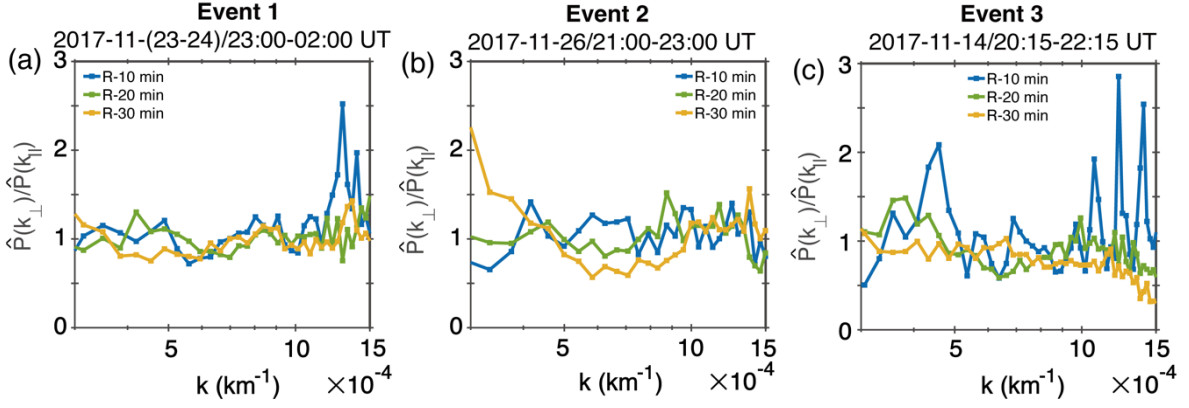

**Figure 9.** The ratio $\hat{P}(k_\perp)/\hat{P}(k_\parallel)$ of fluctuations within the $\hat{k}\hat{b}_0$ plane. $\hat{P}(k_\perp)$ represents the normalized perpendicular wavenumber spectrum, and $\hat{P}(k_\parallel)$ represents the normalized parallel wavenumber spectrum. (a) Event 1: during 23:00-02:00 UT on 2017 November 23-24. (b) Event 2: during 21:00-23:00 UT on 2017 November 26. (c) Event 3: during 20:15-22:15 UT 2017 November 14.

## 4. Further Analysis on MHD Modes

Given the small fluctuations with a strong uniform background magnetic field, the MHD model has three MHD eigenmodes (Alfvén, fast, and slow modes), roughly taking the place of exact nonlinear solutions (Cho & Lazarian 2003). Using the term 'mode' in this study, we refer to the carriers of turbulent fluctuations, which are not dependent on the propagation properties of classical linear waves. The incompressible fluctuations whose displacement vectors are perpendicular to the $\hat{k}\hat{b}_0$ plane are defined as Alfvén modes, whereas fluctuations whose displacement vectors are within the $\hat{k}\hat{b}_0$ plane are defined as magnetosonic modes (slow and fast modes). In this section, we present further analysis on whether fluctuations under such definition can propagate like classical linear waves and satisfy theoretical dispersion relations of MHD modes. We discuss the possible physical explanations of our observations by taking the 30 minute data set of Event 1 as an example.

### 4.1. $\delta B_{\perp 1}$ Fluctuations and Alfvén Modes

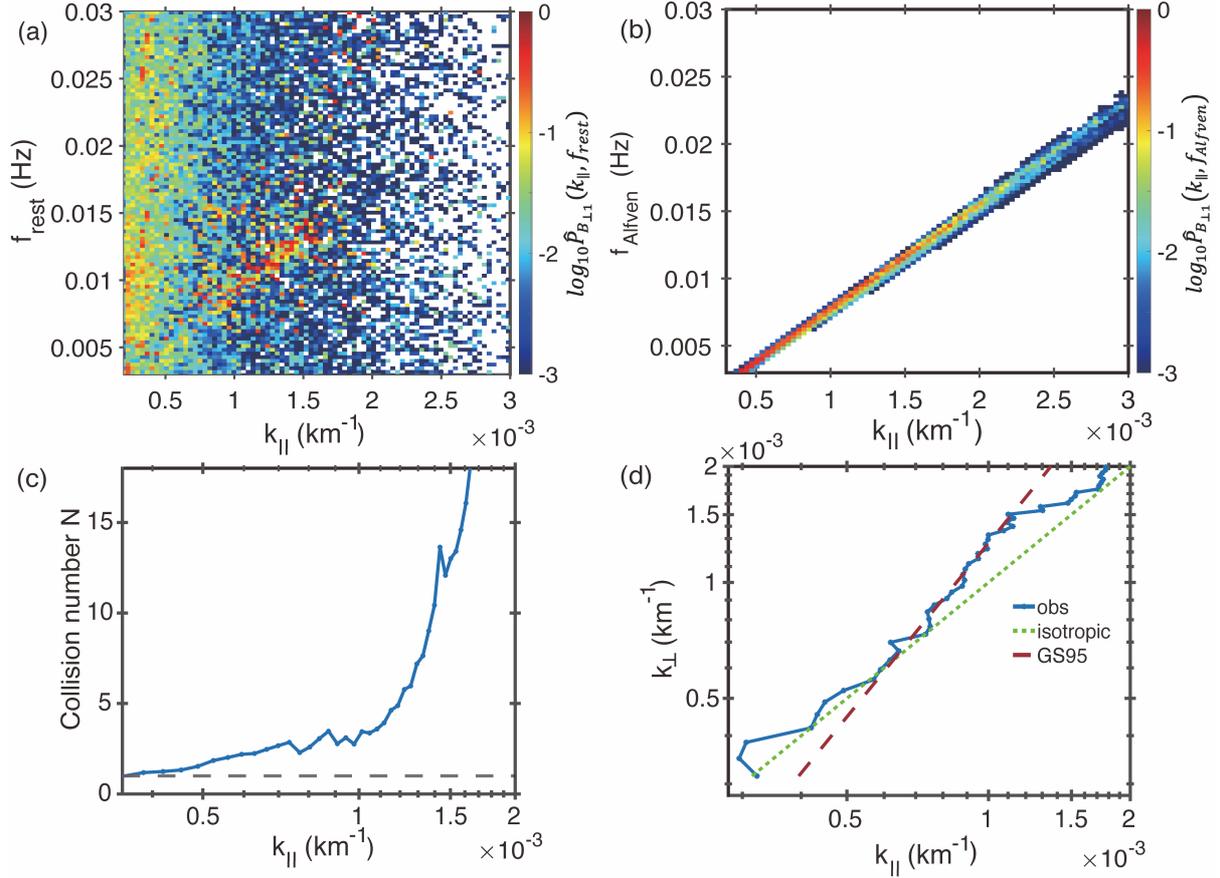

**Figure 10.** (a) The normalized $f_{rest} - k_\parallel$ wavelet power spectra of $\delta B_{\perp 1}$ fluctuations in the rest frame of the solar wind. $\hat{P}_{B_{\perp 1}}(k_\parallel, f_{rest}) = P_{B_{\perp 1}}(k_\parallel, f_{rest})/P_{B_{\perp 1},max}(k_\parallel, f_{rest})$. (b) The normalized theoretical dispersion relation of Alfvén modes. $\hat{P}_{B_{\perp 1}}(k_\parallel, f_{Alfven}) = P_{B_{\perp 1}}(k_\parallel, f_{Alfven})/P_{B_{\perp 1},max}(k_\parallel, f_{Alfven})$. (c) Collision number N versus $k_\parallel$. The horizontal dashed line marks $N = 1$. (d) The variation of $k_\perp$ versus $k_\parallel$ for magnetic power of $\delta B_{\perp 1}$ fluctuations (blue curve). The green dashed line represents isotropy $k_\parallel = k_\perp$. The red dashed line represents the Goldreich-Sridhar scaling $k_\parallel \propto k_\perp^{2/3}$. These figures use the 30-minute dataset of Event 1.

The frequency $f_{rest}$ is obtained by correcting the Doppler shift $f_{rest} = f_{sc} - \frac{\mathbf{k} \cdot \mathbf{V_{sw}}}{2\pi}$, where $f_{rest}$ represents the frequency in the rest frame of the solar wind and $f_{sc}$ represents the frequency in the spacecraft frame. Considering that the SVD method only determines the propagation direction ($\hat{\mathbf{k}}$), we calculate $f_{rest}$ using wavevectors $\mathbf{k}_l$ derived from timing analysis of $\delta B_{\perp 1}$ fluctuations. Here, $\mathbf{k}_l$ is expressed as $\mathbf{k}_{\delta B_{\perp 1}}$. Although we set a stringent criterion $\phi_{\hat{k}k_l} < 10°$ for $\delta B_{\perp 1}$ fluctuations in section 2.3, $\mathbf{k}_{\delta B_{\perp 1}}$ deviates from the direction of minimum variance vectors of magnetic field fluctuations ($\hat{\mathbf{k}}$), resulting in $f_{rest}$ uncertainties between $|f_{sc} - \frac{|\mathbf{k}_{\delta B_{\perp 1}}||V_{sw}|\cos(\theta_{kV_{sw}}-10°)}{2\pi}|$ and $|f_{sc} - \frac{|\mathbf{k}_{\delta B_{\perp 1}}||V_{sw}|\cos(\theta_{kV_{sw}}+10°)}{2\pi}|$. It is challenging to correct $f_{rest}$ uncertainties because $\phi_{\hat{k}k_l}$ varies with the time $t$ and $f_{sc}$. To simplify, we correct $f_{rest}$ uncertainties with uniform angles between $-10°$ and $10°$. The distribution trend will not change when we change the correction angle; thus, our main conclusions are solid.

Figure 10a presents the observed $f_{rest} - k_{\parallel}$ power spectra of $\delta B_{\perp 1}$ in the rest frame of the solar wind after a uniform angle correction ($\phi_{\hat{k}k_l} = 10°$). Similarly to $k_{\perp} - k_{\parallel}$ magnetic power spectra, we construct a set of 100×400 bins to obtain $k_{\parallel} - f_{rest}$ magnetic power spectra. Each bin subtends approximately the same parallel wavenumber and frequency in the plasma flow frame. We sum magnetic power in each bin at all times. To cover all MHD-scale fluctuations, we set the maximum parallel wavenumber as $k_{\parallel,max} = 1.1 \times \frac{0.2}{d_i}$ and the maximum frequency as $f_{rest} = f_{ci}$. The magnetic power $\hat{P}_{B_{\perp 1}}(k_{\parallel}, f_{rest}) = P_{B_{\perp 1}}(k_{\parallel}, f_{rest})/P_{B_{\perp 1},max}(k_{\parallel}, f_{rest})$ is normalized by the maximum power in all bins. The $f_{rest} - k_{\parallel}$ magnetic power spectrum shows two types of fluctuations: (1) No clear relation between the frequency ($f_{rest}$) and parallel wavenumber ($k_{\parallel}$) at $k_{\parallel} < 5 * 10^{-4}\ km^{-1}$. (2) A branch of an apparent linear power enhancement at $5 * 10^{-4} < k_{\parallel} < 1.5 * 10^{-3}\ km^{-1}$. However, Figure 5d shows that the reduced magnetic power spectrum $\hat{P}(k_{\parallel})$ is continuous with respect to $k_{\parallel}$.

Based on the compressible MHD theory, the fluctuations whose polarization is perpendicular to $\hat{k}\hat{b}_0$ plane in wavevector space are expected from Alfvén modes. To further understand the observed magnetic power distributions out of $\hat{k}\hat{b}_0$ plane, we first compare $f_{rest} - k_{\parallel}$ power spectra of $\delta B_{\perp 1}$ with Alfvén-mode theoretical dispersion relations. The theoretical frequencies of Alfvén modes are given by

$$\omega_A^2(k_{\parallel}) = k_{\parallel}^2 V_A^2, \tag{4}$$

where $V_A$ is the Alfvén speed. In Figure 10b, we assume that magnetic power is totally provided by Alfvén modes. The theoretical dispersion relation of Alfvén modes is roughly consistent with the linear branch of the magnetic power spectrum (Figure 10a). This result provides direct observational evidence that these fluctuations may originate from Alfvén modes. It is noteworthy that $f_{rest}$ are slightly higher than theoretical Alfvén frequencies if we do not correct the uncertainties resulting from $\phi_{\hat{k}k_l}$. Comparing observations with the theoretical dispersion relations, we obtain the best match with the angle correction $\phi_{\hat{k}k_l} = 10°$.

To further understand the fluctuations without apparent linear relations between the frequency ($f_{rest}$) and parallel wavenumber ($k_{\parallel}$) at $k_{\parallel} < 5 * 10^{-4}\ km^{-1}$, we calculate the collision number defined as $N = (\frac{\tau_{nl}}{\tau_A})^2 = (\frac{V_A l_{\perp}}{\delta V_l l_{\parallel}})^2$, where $\tau_{nl} = \frac{l_{\perp}}{v_l}$ represents the nonlinear interaction time, $\tau_A = \frac{V_A}{l_{\parallel}}$ represents the linear interaction time, $\delta V_l$ represents perpendicular velocity fluctuations, and $l_{\perp}(l_{\parallel})$ represents perpendicular (parallel) length scale. The strength of the nonlinear effects can be estimated by the collision number. Therefore, turbulence is typically divided into weak ($N \gg 1$) and strong ($N > \sim 1$) turbulence regimes. When we analyze magnetic field fluctuations out of the $\hat{k}\hat{b}_0$ plane ($\delta B_{\perp 1}$), the collision number can be approximately expressed as $N = (\frac{V_A l_{\perp}}{\delta V_l l_{\parallel}})^2 \sim (\frac{B_0 l_{\perp}}{\delta B_{\perp 1} l_{\parallel}})^2$. Using the wavelet technique, we obtain the time-frequency magnetic power spectra and normalize the magnetic power to time points. Then, we construct a set of 100×400 bins to obtain time-normalized magnetic power spectra $\bar{P}_{B_{\perp 1}}(k_{\parallel}, f_{sc})$. Each bin subtends approximately the same parallel wavenumber and spacecraft-frame frequency. To cover all MHD-scale fluctuations, we set the maximum parallel wavenumber $k_{\parallel,max} = 1.1 \times \frac{0.2}{d_i}$ and the maximum spacecraft-frame frequency $f_{sc,max} = f_{ci}$. This study approximately estimates the magnetic field fluctuations by integrating $\int_{f_{sc}}^{\infty} \bar{P}_{B_{\perp 1}}(k_{\parallel}, f_{sc}) df_{sc} \sim f_{sc} \bar{P}_{B_{\perp 1}}(k_{\parallel}, f_{sc})$ over $f_{sc}$ and then $k_{\parallel}$.

Figure 10c shows collision number $N$ versus $k_{\parallel}$. We summarize the comparisons between $f_{rest} - k_{\parallel}$ spectra and collision numbers as follows: (1) At $k_{\parallel} < 5 * 10^{-4}$ $km^{-1}$, the collision number is very close to 1 (Figure 10c), suggesting strong turbulence. At approximately the same wavenumbers, no clear relation between $f_{rest}$ and $k_{\parallel}$ exists (accounting for ~50% of total power; Figure 10a). This is likely because the fast decay of the turbulence within one wave period prevents the Alfvén waves from propagating. (2) At $5 * 10^{-4} < k_{\parallel} < 1.5 * 10^{-3} km^{-1}$, the collision number becomes slightly larger than 1 (Figure 10c), suggesting that turbulence becomes relatively weaker. This may be because some physical dissipation mechanisms diminish the turbulent amplitudes, leading to weaker nonlinear dynamics (Howes et al. 2011). It is worth noting that we deduce that the turbulence is still in the strong regime because the weak turbulence regime needs more collisions ($N \gg 1$) than what we observed. (3) At $k_{\parallel} > 1.5 * 10^{-3} km^{-1}$, the calculated collision number is of no physical meaning due to the limited data samples (Figure 10a).

The recent simulation also presents results similar to the power spectrum in Figure 10a (Gan et al. 2022). Gan et al. proposed an alternate interpretation for the low-frequency nonlinear fluctuations. They believed that the 'non-wave' power does not belong to any eigenmode branches since fluctuations cannot fit any dispersion relations. However, we have reservations about their explanations because there is no reason to completely separate the fluctuations in a continuous reduced magnetic power spectrum $\hat{P}(k_{\parallel})$ (Figure 5d).

In Section 3.1, we have shown that the reduced wavenumber spectra of $\delta B_{\perp 1}$ fluctuations roughly follow the scalings: $\hat{P}(k_{\perp}) \propto k_{\perp}^{-5/3}$ and $\hat{P}(k_{\parallel}) \propto k_{\parallel}^{-2}$, consistent with the Goldreich & Sridhar (1995) theory. To obtain a more intuitive wavenumber relationship, we extract the relation of $k_{\perp}$ versus $k_{\parallel}$ by taking the same values of the magnetic power spectrum at the $k_{\perp}$ and $k_{\parallel}$ axes. Figure 10d shows the variation of $k_{\perp}$ versus $k_{\parallel}$ for magnetic power of $\delta B_{\perp 1}$ fluctuations (blue curve). The green dashed line represents isotropy $k_{\parallel} = k_{\perp}$, and the red dashed line denotes the Goldreich-Sridhar scaling $k_{\parallel} \propto k_{\perp}^{2/3}$. At wavenumber $k_{\parallel} < 1.1 \times 10^{-3} km^{-1}$ ($k_{\perp} < 1.5 \times 10^{-3} km^{-1}$), the observed variation of $k_{\perp}$ versus $k_{\parallel}$ follows the Goldreich-Sridhar scaling $k_{\parallel} \propto k_{\perp}^{\frac{2}{3}}/l_0^{\frac{1}{3}}$, with the normalization consistent with the correlation length $l_0$ obtained in Section 3.1. The collision number $N$ is close to one at approximately the same parallel wavenumber $k_{\parallel} < 1.1 \times 10^{-3} km^{-1}$ (Figure 10c). Therefore, our observations provide direct evidence for the validity of the Goldreich-Sridhar scaling in the solar wind.

### 4.2. Fluctuations within the $\hat{k}\hat{b}_0$ Plane and Magnetosonic Modes

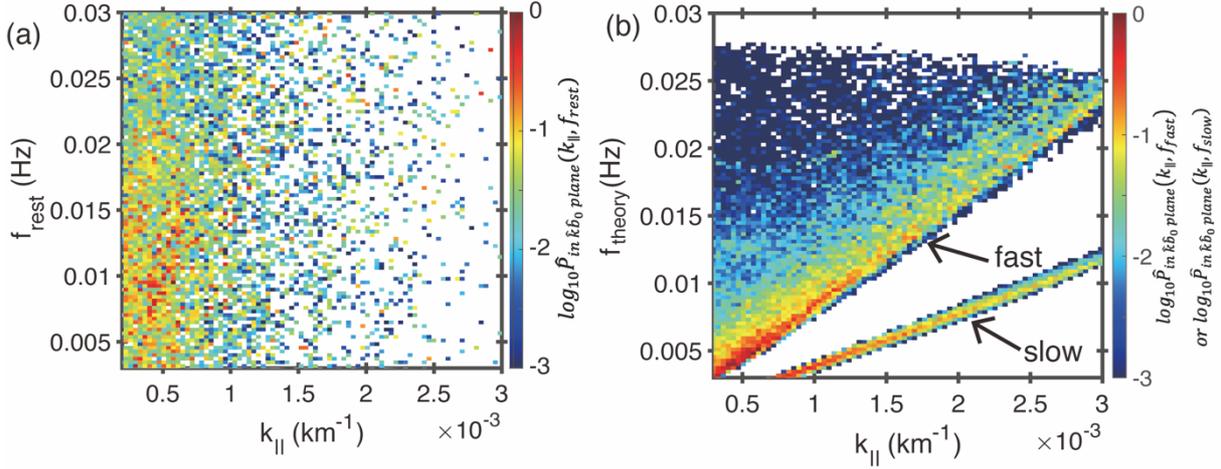

**Figure 11:** (a) Normalized $f_{rest} - k_{\parallel}$ wavelet power spectra of the magnetic field within $\hat{k}\hat{b}_0$ plane in the rest frame of the solar wind. $\hat{P}_{in\ \hat{k}\hat{b}_0\ plane}(k_{\parallel}, f_{rest}) = P_{in\ \hat{k}\hat{b}_0\ plane}(k_{\parallel}, f_{rest})/P_{in\ \hat{k}\hat{b}_0\ plane,max}(k_{\parallel}, f_{rest})$. (b) The normalized theoretical dispersion relation of fast and slow modes, where $\hat{P}_{in\ \hat{k}\hat{b}_0\ plane}(k_{\parallel}, f_{fast}) = P_{in\ \hat{k}\hat{b}_0\ plane}(k_{\parallel}, f_{fast})/P_{in\ \hat{k}\hat{b}_0\ plane,max}(k_{\parallel}, f_{fast})$ and $\hat{P}_{in\ \hat{k}\hat{b}_0\ plane}(k_{\parallel}, f_{slow}) = P_{in\ \hat{k}\hat{b}_0\ plane}(k_{\parallel}, f_{slow})/P_{in\ \hat{k}\hat{b}_0\ plane,max}(k_{\parallel}, f_{slow})$. These figures use the 30-minute dataset of Event 1.

Given that the magnetic power within the $\hat{k}\hat{b}_0$ plane only plays a limited role, the reliability of quantitative analysis is questionable. Besides, it is more difficult to correct the $f_{rest}$ uncertainties resulting from $\phi_{kk_l}$, because we set a more relaxed criterion $\phi_{kk_l} < 30°$ for in-plane fluctuations in order to obtain enough samplings. Therefore, magnetic field fluctuations within the $\hat{k}\hat{b}_0$ plane are discussed qualitatively. Figure 11a shows $f_{rest} - k_{\parallel}$ power spectra of the magnetic field within the $\hat{k}\hat{b}_0$ plane in the rest frame of the solar wind without any angle correction for $f_{rest}$. The in-plane magnetic power $\hat{P}_{in\ \hat{k}\hat{b}_0\ plane}(k_{\parallel}, f_{rest}) = P_{in\ \hat{k}\hat{b}_0\ plane}(k_{\parallel}, f_{rest})/P_{in\ \hat{k}\hat{b}_0\ plane,max}(k_{\parallel}, f_{rest})$ is normalized by the maximum power in all bins. The magnetic power is concentrated in $k_{\parallel} < 1 \times 10^{-3}$ $km^{-1}$ and shows no relationship between $f_{rest}$ and $k_{\parallel}$.

According to the ideal MHD theory, magnetic field fluctuations within the $\hat{k}\hat{b}_0$ plane are most likely provided by compressible magnetosonic modes (fast and slow modes). The theoretical frequencies of fast and slow modes are given by

$$\omega_{F,S}^2(k) = k^2 V_A^2 \left[ \frac{1 + \beta_p}{2} \pm \sqrt{\frac{(1 + \beta_p)^2}{4} - \beta_p (\frac{k_{\parallel}}{k})^2} \right], \tag{5}$$

where $k = \sqrt{k_{\parallel}^2 + k_{\perp}^2}$ is the wavenumber, and $k_{\parallel}$ is the parallel wavenumber to $\boldsymbol{B}_0$. In Figure 11b, we assume that magnetic power is totally provided by either fast or slow modes. Since $\theta_{kB_0}$ between $\hat{k}$ and $\hat{b}_0$ varies with wavenumbers, fast modes do not show linear dispersion relations (Figure 11b). Given relatively large $f_{rest}$ uncertainties, we only focus on the parallel wavenumber distributions of the magnetic power within the $\hat{k}\hat{b}_0$ plane. If only fast modes existed, fast-mode magnetic power would roughly cover the wavenumber distributions of the observed magnetic power ($k_\parallel < 1 \times 10^{-3} \ km^{-1}$) in Figure 11a. However, if there are only slow modes within the $\hat{k}\hat{b}_0$ plane, the magnetic power would be concentrated in larger parallel wavenumbers ($k_\parallel > 7 \times 10^{-4} \ km^{-1}$) than actual observations. Overall, it is challenging to identify mode compositions of the fluctuations by comparing the disordered distributions with theoretical dispersion relations.

In low-$\beta_p$ plasma, magnetic field fluctuations are expressed as $(\frac{\delta B}{B_0})_{slow} \sim \frac{a}{V_A^2} \delta V_{slow}$ for slow modes and $(\frac{\delta B}{B_0})_{fast} \sim \frac{\delta V_{fast}}{V_A}$ for fast mode, where $a$ is the sound speed, and $V_A$ is the Alfvén speed. $\delta V_{slow}$ and $\delta V_{fast}$ represent the fluctuating velocity of slow and fast modes, respectively (Cho & Lazarian 2003). If $\delta V_{slow} \sim \delta V_{fast}$, we obtain $(\frac{\delta B}{B_0})_{slow} = \sqrt{\beta_p} (\frac{\delta B}{B_0})_{fast}$. Therefore, fast modes theoretically may provide more magnetic field fluctuations than slow modes when $\beta_p < 1$ (Zhao et al. 2021). For these three events, the proton plasma $\beta_p$ is $\sim 0.3$, in accord with this condition. Moreover, our observations show that magnetic field fluctuations within the $\hat{k}\hat{b}_0$ plane present isotropic behaviors and follow scalings $\hat{P}(k_\perp) \propto k_\perp^{-3/2}$ and $\hat{P}(k_\parallel) \propto k_\parallel^{-3/2}$, consistent with fast modes (e.g., Cho & Lazarian 2002; Makwana & Yan 2020). Therefore, we deduce indirectly that magnetic field fluctuations within the $\hat{k}\hat{b}_0$ plane more likely originate from fast modes. Quantitative analysis for compressible fluctuations will be the subject of our future studies.

## 5. Summary

This study presents observations of three-dimensional magnetic power spectra in wavevector space to investigate the anisotropy and scalings of sub-Alfvénic solar wind turbulence at the MHD scale using the MMS spacecraft. The magnetic power spectra are organized in a new coordinate determined by $\hat{k}$ and $\hat{b}_0$ in Fourier space, as described in Section 2.2. This study utilizes two approaches to determine wavevectors: the singular value decomposition method and multi-spacecraft timing analysis. The combination of the two methods allows an examination of the properties of magnetic field fluctuations in terms of mode compositions independent of any spatiotemporal hypothesis.

The specifics of our findings are summarized below.

1. The magnetic power spectra of $\delta B_{\perp 1}$ (in the direction perpendicular to $\hat{k}$ and $\hat{b}_0$) are prominently stretched along the $k_\perp$ axis, indicating a faster cascade in the perpendicular direction. Moreover, such anisotropy increases as the wavenumber increases. The reduced power spectra of $\delta B_{\perp 1}$ fluctuations follow the Goldreich-Sridhar scalings: $\hat{P}(k_\perp) \propto k_\perp^{-5/3}$ and $\hat{P}(k_\parallel) \propto k_\parallel^{-2}$. $\delta B_{\perp 1}$ fluctuations are more anisotropic using a shorter-interval average magnetic field, suggesting fluctuations are more likely aligned with the *local* magnetic field than a global magnetic field.

2. The magnetic power spectra within the $\hat{k}\hat{b}_0$ plane show isotropic behaviors: the perpendicular power distributions are roughly comparable to parallel distributions. The reduced magnetic power spectra within the $\hat{k}\hat{b}_0$ plane follow the scalings $\tilde{P}(k_\perp) \propto k_\perp^{-3/2}$ and $\tilde{P}(k_\parallel) \propto k_\parallel^{-3/2}$.

3. Comparing observational frequency-wavevector spectra in the rest frame of the solar wind with theoretical dispersion relations of MHD modes, we find that $\delta B_{\perp\perp}$ fluctuations are consistent with Alfvén modes. In contrast, we deduce that the magnetic field fluctuations within the $\hat{k}\hat{b}_0$ plane more likely originate from fast modes based on their isotropic behaviors and scalings.

4. We estimated the number (N) of wave packet collisions needed to induce the turbulence cascade. We provide direct observational evidence for the scale-dependent anisotropy from the Goldreich-Sridhar model in the scale range where the critical balance (N≥1) holds.

## Acknowledgments

We would like to thank the members of the MMS spacecraft team, NASA's Coordinated Data Analysis Web, and NASA OMNIWeb. The MMS data and OMNI data are available at https://cdaweb.gsfc.nasa.gov. Data analysis was performed using the IRFU-MATLAB analysis package available at https://github.com/irfu/irfu-matlab and the SPEDAS analysis software available at http://themis.ssl.berkeley.edu.

Table 1

Sub-Afvenic fluctuations in solar wind turbulence

| No. | Start Time (UT) | End Time (UT) | Scale (minutes) | Location ($R_E$) | $\beta_p$ | $V_{sw}$ ($km\,s^{-1}$) | $f_{ci}$ (Hz) | $r_{ci}$ (km) | $d_i$ (km) | $\langle \frac{\delta B_{rms}}{|B|} \rangle_{30min}$ | $k_\perp$ and $k_\parallel$ exponent for $\hat{P}_{\delta B_{\perp 1}}$ | $k_\perp$ and $k_\parallel$ exponent for $\hat{P}_{\delta B_{\perp 2}}$ |
|---|---|---|---|---|---|---|---|---|---|---|---|---|
| 1 | 2017-11-23/23:00 | 2017-11-24/02:00 | 180 | [17, 18, 6] | 0.32 | 385 | 0.1 | 31.8 | 80.6 | $0.28 \pm 0.02$ | $\hat{P}(k_\perp) \propto k_\perp^{-1.66}$ $\hat{P}(k_\parallel) \propto k_\parallel^{-2.06}$ | $\hat{P}(k_\perp) \propto k_\perp^{-1.51}$ $\hat{P}(k_\parallel) \propto k_\parallel^{-1.53}$ |
| 2 | 2017-11-26/21:00 | 2017-11-26/23:00 | 120 | [17, 17, 6] | 0.28 | 336 | 0.11 | 26.7 | 71.7 | $0.29 \pm 0.05$ | $\hat{P}(k_\perp) \propto k_\perp^{-1.62}$ $\hat{P}(k_\parallel) \propto k_\parallel^{-2.69}$ | $\hat{P}(k_\perp) \propto k_\perp^{-1.5}$ $\hat{P}(k_\parallel) \propto k_\parallel^{-1.51}$ |
| 3 | 2017-11-14/20:15 | 2017-11-14/22:15 | 120 | [17, 17, 6] | 0.37 | 387 | 0.08 | 36.9 | 87.2 | $0.35 \pm 0.14$ | $\hat{P}(k_\perp) \propto k_\perp^{-1.7}$ $\hat{P}(k_\parallel) \propto k_\parallel^{-1.97}$ | $\hat{P}(k_\perp) \propto k_\perp^{-1.47}$ $\hat{P}(k_\parallel) \propto k_\parallel^{-1.47}$ |